\documentclass[aps,prl,twocolumn,showpacs]{revtex4}

\usepackage{graphicx}
\begin{document}


\title{Superconductivity in Rh$_2$Ga$_9$ and Ir$_2$Ga$_9$ without Inversion Symmetry}

\author{T. Shibayama$^{1}$, M. Nohara$^{1,2}$, H. Aruga Katori$^{2,3}$, Y. Okamoto$^{4}$, Z. Hiroi$^{4}$, and H. Takagi$^{1,2,3}$
}
\affiliation{
$^{1}$Department of Advanced Materials, University of Tokyo, Kashiwa, Chiba 277-8561\\
$^{2}$CREST, Japan Science and Technology Agency (JST)\\
$^{3}$RIKEN (The Institute of Physical and Chemical Research), Wako, Saitama 351-0198\\
$^{4}$Institute for Solid State Physics, University of Tokyo, Kashiwa, Chiba 277-8581
}

\date{\today}

\begin{abstract}
Superconductivity with $T_{\rm c}$ $\simeq$ 2 K was discovered in the intermetallic binary compounds Rh$_2$Ga$_9$ and Ir$_2$Ga$_9$. This is the first observation of superconductivity in the Rh-Ga and Ir-Ga binary systems. Both compounds crystallize in a distorted Co$_2$Al$_9$-type structure (monoclinic, space group: $Pc$), which lacks spatial inversion symmetry. 
Specific heat measurements revealed that both compounds are weak-coupling BCS superconductors having an isotropic superconducting gap. Measurements in magnetic fields indicated type-I superconductivity with a critical field $H_{\rm c}(0)$ $\simeq$ 130 Oe for Rh$_2$Ga$_9$ and type-II superconductivity with an upper critical field $H_{\rm c2}(0)$ $\simeq$ 250 Oe for Ir$_2$Ga$_9$.
\end{abstract}

\maketitle

Superconductors without spatial inversion symmetry have been attracting considerable interest. In such systems, if the antisymmetric spin-orbit coupling (SOC) is strong enough, a conventional classification of the pair wave function $s$-, $p$- or $d$-wave for the orbital part and singlet or triplet for the spin part is not valid anymore, and unconventional superconductivity with a nontrivial pair wave function is expected to appear \cite{rf:CePt3Si,rf:Frigeri-1,rf:Frigeri-2}. 
Until now, superconductors without spacial inversion symmetry, and presumably having large SOC, have been discovered in compounds that consist of heavy transition metal (5$d$), lanthanoid (4$f$) and actinoid (5$f$) elements. Among them, CePt$_3$Si \cite{rf:CePt3Si}, UIr \cite{rf:UIr}, CeRhSi$_3$ \cite{rf:CeRhSi3} and CeIrSi$_3$ \cite{rf:CeIrSi3} constitute a family of heavy-fermion superconductors, and the nontrivial symmetry of Cooper pairs has been discussed in relation to the lack of inversion symmetry. Superconductivity in these systems, however, is located at the critical vicinity to the magnetic quantum critical point and coexists with antiferromagnetism or ferromagnetism. This can potentially make these compounds more fascinating; however, it may make the playground too complicated to capture the physics of inversion symmetry breaking and superconductivity. 

In contrast, transition metal compounds with electron-phonon-mediated superconductivity give us an opportunity to investigate the bare effects of inversion symmetry breaking due to much weaker electron correlation for 4$d$ and 5$d$ systems. In addition, the magnitude of the SOC can be tuned by utilizing 4$d$ (small SOC) and 5$d$ (large SOC) elements in isoelectronic and isostructural compounds. 
Such an interesting case might be realized in the transition metal borides Li$_2$Pd$_3$B \cite{rf:Li2Pd3B} and Li$_2$Pt$_3$B \cite{rf:Li2Pt3B} in which the SOC for Pt (5$d$) is much larger than that for Pd (4$d$). 
The penetration depth \cite{rf:Li2Pt3B-pene} and $^{11}$B Knight shift \cite{rf:Li2Pt3B-nmr} have suggested unconventional superconductivity with line nodes and significant spin-triplet component in pair wave function for Li$_2$Pt$_3$B, while conventional superconductivity with an isotropic gap for Li$_2$Pd$_3$B.

\begin{figure}
\center
\includegraphics[width=7cm]{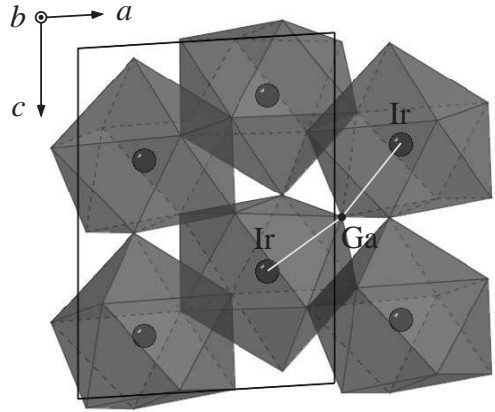}
\caption{\label{fig:structure}
Crystal structure of Rh$_2$Ga$_9$ and Ir$_2$Ga$_9$.
Monocapped square antiprisms [IrGa$_9$] with different orientations relative to [001] are stacked along [001] direction. The centering Ir atoms are represented by gray circles. The solid lines represent a monoclinic unit cell. White lines indicate a characteristic Ir-Ga-Ir bond, which determines the magnitude of the broken inversion symmetry (see text).
}
\end{figure}

This motivated us to explore new superconductors with $4d$ and $5d$ elements without spatial inversion symmetry. 
During the course of this study, we discovered superconductivity at about 2 K in the binary intermetallic compounds Rh$_2$Ga$_9$ and Ir$_2$Ga$_9$. This is the first report on superconducting binary gallides containing Rh and Ir. 
In this Letter, we reveal their superconducting and normal state properties.

Rh$_2$Ga$_9$ and Ir$_2$Ga$_9$ crystallize in a monoclinic (distorted Co$_2$Al$_9$-type) structure with the space group $Pc$ (No. 7), as shown in Fig. \ref{fig:structure} \cite{rf:structure}. The structure is characterized by a monocapped square antiprism centered at Ir (Rh). An Ir(Rh)-Ga-Ir(Rh) bond angle of 165.8$^{\circ}$ (164.5$^{\circ}$) for Ir$_2$Ga$_9$ (Rh$_2$Ga$_9$) is much smaller than 180$^{\circ}$ and comparable with the Pt-B-Pt bond angle of 150.1$^{\circ}$ for antiperoviskite Li$_2$Pt$_3$B. This ensures significant inversion symmetry breaking in Rh$_2$Ga$_9$ and Ir$_2$Ga$_9$. Rhodium ($4d$) and particularly iridium ($5d$) are heavy elements, and the SOC should be invoked as an important ingredient for low-energy electronic states.

Polycrystalline samples were prepared by argon arc melting and subsequent heat treatment at 500 $^{\circ}$C under vacuum for one week. Powder X-ray diffraction measurements revealed the formation of Rh$_2$Ga$_9$ and Ir$_2$Ga$_9$ without any noticeable impurity phases. The estimated lattice parameters were almost the same as reported in Ref. \cite{rf:structure}.
A low residual resistivity of $\rho_{\rm 0}$ $\simeq$ 1 $\mu\Omega$cm and a large residual resistivity ratio (RRR) of $\sim$ 150 for both Rh$_2$Ga$_9$ and Ir$_2$Ga$_9$ suggest high quality of the samples.
Magnetic, transport and thermal measurements were conducted by using Magnetic Property Measurement System (MPMS, Quantum Design), Physical Property Measurement System (PPMS, Quantum Design) and $^3$He refrigerator (Heliox, Oxford).

\begin{figure}
\center
\includegraphics[width=7cm]{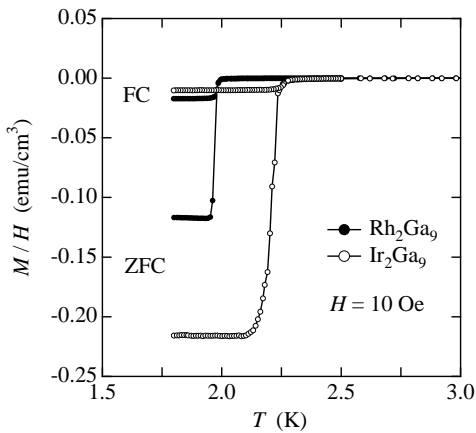}
\caption{\label{fig:M}
Temperature dependence of the magnetization of Rh$_2$Ga$_9$ and Ir$_2$Ga$_9$. 
The measurements were conducted in an applied field of $H$ = 10 Oe with zero-field-cooled (ZFC) and field-cooled (FC) processes. 
}
\end{figure}

\begin{figure}
\center
\includegraphics[width=8cm]{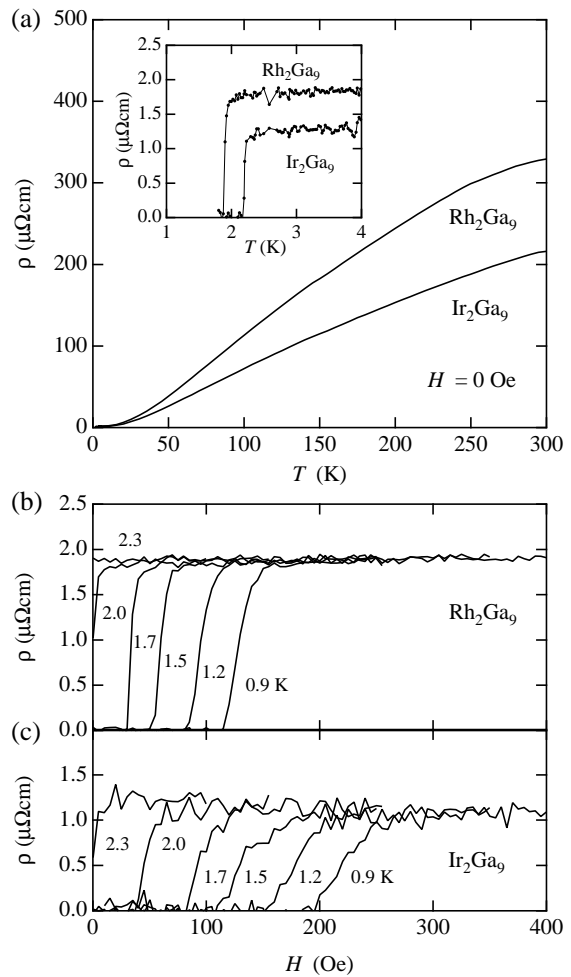}
\caption{\label{fig:rho}
(a) Temperature dependence of electrical resistivity $\rho$ of Rh$_2$Ga$_9$ and Ir$_2$Ga$_9$ in zero applied field. (b) and (c) Low-temperature resistivity as a function of magnetic field $H$. The measurements were conducted on decreasing $H$  ($\parallel$ $j$ : current) from the normal state at constant temperatures (0.9, 1.2, 1.5, 1.7, 2.0 and 2.3 K) with a sufficiently low current density of $\sim$ 0.5 A/cm$^2$. 
}
\end{figure}

The evidence for superconductivity in Rh$_2$Ga$_9$ and Ir$_2$Ga$_9$ was found in the magnetization $M(T)$ and electrical resistivity $\rho(T)$, as shown in Figs. \ref{fig:M} and \ref{fig:rho}(a), respectively. 
A large Meissner signal was clearly observed in the $M(T)$ curve below $T_{\rm c}$ =  1.9 and 2.2 K for Rh$_2$Ga$_9$ and Ir$_2$Ga$_9$, respectively.
Simultaneously, the $\rho(T)$ exhibited a zero-resistive state. The field-cooled (FC) magnetization reached $\sim$ 20\% of the perfect diamagnetism at low temperatures. This large Meissner effect is the hallmark of bulk superconductivity.

\begin{figure}
\center
\includegraphics[width=9cm]{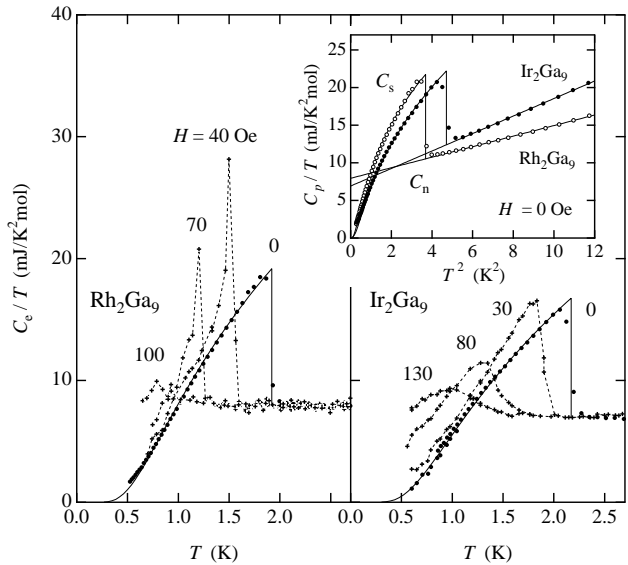}
\caption{\label{fig:C}
Electronic specific heat divided by temperature $C_{\rm e}/T$ as a function of temperature $T$ in various magnetic fields for Rh$_2$Ga$_9$ and Ir$_2$Ga$_9$. 
Solid lines for the $H$ = 0 data represent $C_{\rm e}/T$ expected from the weak-coupling BCS theory \cite{rf:BCS}.
The inset shows specific heat divided by temperature $C_{p}/T$ as a function of $T^2$ in zero applied field. The solid lines represent a fit to $C_{\rm n}/T$ $=$ $\gamma_{\rm n} + \beta T^2$. 
}
\end{figure}

Further support for bulk superconductivity was obtained from the specific heat $C_{p}(T)$, where a clear jump at the superconducting transition was observed, as shown in Fig. \ref{fig:C}.
In order to accurately determine bulk $T_{\rm c}$ in zero magnetic field, an idealized jump at $T_{\rm c}$ was assumed to satisfy the entropy conservation at the transition. This yielded an estimate of $T_{\rm c}$ = 1.9 and 2.2 K and $\Delta C_{p}/T_{\rm c}$ = 11.3 and 9.8 mJ/K$^2$mol for Rh$_2$Ga$_9$ and Ir$_2$Ga$_9$, respectively. 
A standard analysis yielded the normal-state $T$-linear specific heat coefficient $\gamma_{\rm n}$ = 7.9 and 6.9 mJ/K$^2$mol and Debye temperature $\Theta_{\rm D}$ = 312 and 264 K for Rh$_2$Ga$_9$ and Ir$_2$Ga$_9$, respectively. 
By using these values, we estimated  $\Delta C_{p}/\gamma_{\rm n}T_{\rm c}$ $\simeq$ 1.4 for both Rh$_2$Ga$_9$ and Ir$_2$Ga$_9$, which is almost identical to the value expected from the BCS weak-coupling limit ($\Delta C_{p}/\gamma_{\rm n}T_{\rm c}$ $=$ 1.43). 
In zero applied field, $C_{p}(T)$ showed exponential temperature dependence at low temperatures. Indeed, the $C_{p}(T)$ data can be fitted reasonably by those expected from the weak-coupling BCS theory (represented by solid lines in Fig. \ref{fig:C}) \cite{rf:BCS}.
All of these results suggest that both Rh$_2$Ga$_9$ and Ir$_2$Ga$_9$ are weak-coupling superconductors with an isotropic superconducting gap.

\begin{figure}
\center
\includegraphics[width=9.1cm]{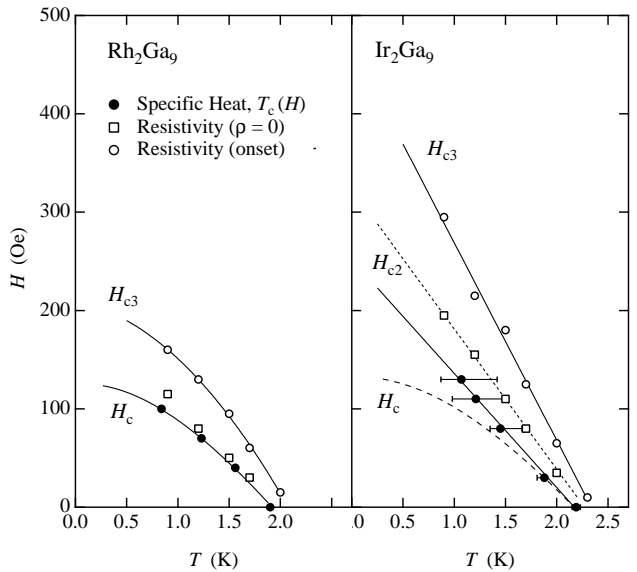}
\caption{\label{fig:H-T}
Magnetic field ($H$) versus temperature ($T$) phase diagram of Rh$_2$Ga$_9$ and Ir$_2$Ga$_9$. Filled circles and open squares represent the superconducting transition temperatures determined by specific heat and resistivity, respectively. Open squares represent the onset magnetic field of surface superconductivity. Thermodynamic critical field $H_{\rm c}(T)$, determined from the zero-field specific heat, is shown by solid and broken lines for Rh$_2$Ga$_9$ and Ir$_2$Ga$_9$, respectively. 
}
\end{figure}

Although the zero-field specific heat data were similar, the behavior in magnetic fields was distinctly different between Rh$_2$Ga$_9$ and Ir$_2$Ga$_9$. This can be clearly seen in Fig. \ref{fig:C}. 
Rh$_2$Ga$_9$ exhibited a divergent $C_{p}(T)$ anomaly at the transition, the characteristic of a first-order transition,  when $H$ $\neq$ 0. This strongly suggests that Rh$_2$Ga$_9$ is a type-I superconductor and that the superconducting transition becomes a first-order transition in magnetic fields. 
As shown in Fig. \ref{fig:H-T}, the $H_{\rm c}$ -  $T_{\rm c}$ curve, obtained from the specific heat anomaly, agreed very well with the thermodynamic critical field $H_{\rm c}(T)$, which was estimated from the free-energy difference between the normal state and the superconducting state in zero applied field: $\Delta F(T)$ $=$ $F_{\rm n} - F_{\rm s}$ $=$ $H_{\rm c}^{2}(T)/8\pi$ $=$ $\int_{T}^{T_{\rm c}} \int_{T'}^{T_{\rm c}}(C_{\rm n}/T'' - C_{\rm s}/T'')dT''dT'$. Here, we used the zero-field specific heat data for $C_{\rm s}$ and assumed $C_{\rm n}$ $=$ $\gamma_{\rm n}T$ $+$ $\beta T^3$ (see the inset in Fig. \ref{fig:C}). 
This, together with the first-order transition in $H$, indicates that Rh$_2$Ga$_9$ is a type-I superconductor with a critical field of $H_{\rm c}(T=0)$ $\simeq$ 130 Oe.
In contrast, the specific heat of Ir$_2$Ga$_9$ exhibited a second-order behavior at $T_{\rm c}$ in $H$. The observed $H_{\rm c2}$ - $T_{\rm c}$ curve is located at a higher field than the thermodynamic critical field $H_{\rm c}(T)$, suggesting type-II superconductivity in Ir$_2$Ga$_9$. We estimate a coherence length of $\xi$ $\sim$ 1000 \AA~ from the linearly extrapolated upper critical field $H_{\rm c2}(0)$ $\simeq$ 250 Oe.
Frigeri {\it et al.} predicted that if the spin-triplet component is present due to an antisymmetric SOC, the paramagnetic limiting field $H_{\rm P}$ is enhanced from the value without the SOC, $H_{\rm P}^0$ $\simeq$ $1.85T_{\rm c}$ (in Tesla)~\cite{rf:Frigeri-1,rf:Frigeri-2}. 
However, because of the long coherence length and the resultant much lower orbital limiting field ($\simeq$ 250 Oe) than $H_{\rm P}^0$  $\simeq$ 3.7 T (expected from $T_{\rm c}$ = 2.2 K for Ir$_2$Ga$_9$), we were not able to examine if $H_{\rm P}$ is enhanced and the effect of the antisymmetric SOC is noticeably large in Ir$_2$Ga$_9$.

The long coherence length $\xi$ of the present compounds gives rise to surface superconductivity: when we have a flat plate of a superconductor, with the applied magnetic field parallel to the flat surfaces, superconductivity can exist in the surface layers with a thickness of coherence length, while the bulk region inside of the sample has a zero order parameter \cite{rf:Parks}. 
Figures \ref{fig:rho}(b) and (c) show the resistivity $\rho$ as a function of $H$ for Rh$_2$Ga$_9$ and Ir$_2$Ga$_9$, respectively. The magnetic field $H$ was applied parallel to the current direction and sample surfaces with a parallelepiped shape.
Zero resistivity persisted in the bulk superconducting state up to the critical field $H_{\rm c}(T)$ for type-I Rh$_2$Ga$_9$ and the upper critical field $H_{\rm c2}(T)$ for type-II Ir$_2$Rh$_9$. Even above $H_{\rm c}(T)$ or $H_{\rm c2}(T)$, $\rho(H)$ was still smaller than the normal-state value and gradually increased until $\rho(H)$ reached to the value in the normal state at the surface critical field $H_{\rm c3}(T)$. 
The estimated $H_{\rm c3}(T)$, together with $H_{\rm c}(T)$ and $H_{\rm c2}(H)$, are shown in Fig. \ref{fig:H-T}. By using the relation $H_{\rm c3}$ $=$ $1.7\sqrt{2}\kappa H_{\rm c}$ \cite{rf:Parks}, we estimate a Ginzburg-Landau parameter $\kappa$ $\simeq$ 0.7 and 1.1 for Rh$_2$Ga$_9$ and Ir$_2$Ga$_9$, respectively. 
The latter agrees reasonably with $\kappa$ $=$ $H_{\rm c2}/\sqrt{2}H_{\rm c}$ $\sim$ 1.3. 
These $\kappa$ values indicate that Rh$_2$Ga$_9$ and Ir$_2$Ga$_9$ are located near the boundary of type-I and type-II superconductivity with $\kappa$ $=$ $1/\sqrt{2}$.

The observed electronic specific heat coefficient $\gamma_{\rm n}$ = 6.9 mJ/K$^2$mol for Ir$_2$Ga$_9$ is only slightly enhanced as compared to that obtained from the band calculation $\gamma_{\rm band}$ = 5.9 mJ/K$^2$mol \cite{rf:structure}. This yields an electron-phonon coupling constant of $\lambda_{\rm ep}$ $\sim$ 0.17 assuming $\gamma_{\rm n}$ $=$ $(1+\lambda_{\rm ep})\gamma_{\rm band}$, which is consistent with the weak-coupling limit inferred from the specific heat jump $\Delta C_{p}/\gamma_{\rm n}T_{\rm c}$. We believe that the same is true for Rh$_2$Ga$_9$. 
The rather small enhancement of $\gamma_{\rm n}$ as compared to $\gamma_{\rm band}$ probably indicates that the correlation effect is not significant.
The $\gamma_{\rm n}$ value of 7 $-$ 8 mJ/K$^2$mol at first glance does not appear to be so small. The band calculation indicates that the electronic state at the Fermi level $E_{\rm F}$ is primarily of the Ga $4s$ and $4p$ character \cite{rf:structure}. It is, therefore, more practical to rewrite $\gamma_{\rm n}$ as $\simeq$ 0.84 mJ/K$^2$mol-Ga for Rh$_2$Ga$_9$ and $\simeq$ 0.77 mJ/K$^2$mol-Ga for Ir$_2$Ga$_9$. It is clear that these systems have a low electronic density of states at $E_{\rm F}$. In accord with this, the magnetic susceptibility of these compounds in the normal state was diamagnetic, $\chi$ $\simeq$ $-$ 2.7 $\times$ $10^{-4}$ emu/mol for Rh$_2$Ga$_9$ and $\simeq$ $-$ 3.0 $\times$ $10^{-4}$ emu/mol for Ir$_2$Ga$_9$. 
The small contribution from Rh and Ir to the electronic states near $E_{\rm F}$, as inferred from the band calculation of Ir$_2$Ga$_9$, partly explains the absence of a noticeable SOC effect in both Rh$_2$Ga$_9$ and Ir$_2$Ga$_9$ and the conventional behavior of superconductivity. 
Another point to be noted is that even for Ir$_2$Ga$_9$, the strength of the SOC $\alpha_{\rm SO}$ $\sim$ 200 meV is much smaller than the Fermi energy $E_{\rm F}$ ($\sim$ 12 eV)  \cite{rf:structure}. 
Frigeri {\it et al.} predicted that if $E_{\rm F}$ $\gg$ $\alpha_{\rm SO}$, the mixing of the spin-singlet and spin-triplet states is negligibly small and $T_{\rm c}$ of the spin-singlet state is essentially unchanged by introducing the antisymmetric SOC \cite{rf:Frigeri-1,rf:Frigeri-2}.  
Thus, the almost same $T_{\rm c}$ of Rh$_2$Ga$_9$ and Ir$_2$Ga$_9$ suggests a dominant singlet component in the pair wave function of Ir$_2$Ga$_9$. 
In contrast with Ir$_2$Ga$_9$, Li$_2$Pt$_3$B has the conduction band with a predominantly Pt $5d$ character and a clear decrease in $T_{\rm c}$ as compared to Li$_2$Pd$_3$B, as well as the signature of unconventional superconductivity, was observed. 
It is still not clear, however, whether $\alpha_{\rm SO}$ can be of the order of $E_{\rm F}$ in Li$_2$Pt$_3$B.

\begin{table}[tb]
\begin{center}
\caption{Superconducting and normal state parameters for Rh$_2$Ga$_9$ and Ir$_2$Ga$_9$.} \label{tab:summary}
\begin{tabular}{lrr}
\\
\hline
& Rh$_2$Ga$_9$ & Ir$_2$Ga$_9$ \\
\hline
Transition temperature, $T_{\rm c}$ (K) & 1.9 &  2.2 \\
$T$-linear coefficient, $\gamma_{\rm n}$ (mJ/K$^2$mol)  &  7.9 &  6.9 \\
Debye temperature, $\Theta_{\rm D}$ (K) &   312 &  264 \\
$\Delta C_{p}/\gamma_{\rm n} T_{\rm c}$ &   1.4 &  1.4 \\
Thermodynamic critical field, $H_{\rm c}(0)$ (Oe) &   126 &  133 \\
Critical field, $H_{\rm c}(0)$ (Oe) & $\simeq$ 130 & n.a.  \\
Upper critical field, $H_{\rm c2}$ (Oe) & n.a. & $\simeq$ 250 \\
Ginzburg-Landau parameter, $\kappa$ & $\simeq$ 0.7 & $\simeq$ 1.1\\
\hline
\end{tabular}
\end{center}
\end{table}

In conclusion, by exploring group 9 transition metals (Co, Rh and Ir) and Ga binary systems, we discovered new superconductors Rh$_2$Ga$_9$ and Ir$_2$Ga$_9$ with $T_{\rm c}$ $=$ 1.9 and 2.2 K, respectively. Rh$_2$Ga$_9$ and Ir$_2$Ga$_9$ are the first examples of superconductors in the Rh-Ga and Ir-Ga binary systems. The superconducting and normal state parameters, as summarized in Table \ref{tab:summary},  revealed that Rh$_2$Ga$_9$ and Ir$_2$Ga$_9$ are weak-coupling BCS superconductors with an isotropic superconducting gap. 

This work was partly supported by a Grant-in-Aid for Scientific Research from the Ministry of Education, Culture, Sports, Science and Technology of Japan.

{\it Note added. $-$} Wakui {\it et al.} \cite{rf:Wakui} have recently reported superconductivity in the same compounds \cite{rf:shiba}.

\end{document}